\documentclass[aps,numerical,graphicx,showpacs,reprint,prl]{revtex4-1}
\usepackage{amsmath}
\usepackage{graphicx}

\begin{document}

\def\Journal#1#2#3#4{{#1 }{\bf #2, }{ #3 }{ (#4)}}

\def\BiJ{ Biophys. J.}
\def\Bios{ Biosensors and Bioelectronics}
\def\LNC{ Lett. Nuovo Cimento}
\def\JCP{ J. Chem. Phys.}
\def\JAP{ J. Appl. Phys.}
\def\JMB{ J. Mol. Biol.}
\def\JPC{ J. Phys: Condens. Matter}
\def\CMP{ Comm. Math. Phys.}
\def\LMP{ Lett. Math. Phys.}
\def\NLE{{ Nature Lett.}}
\def\NPB{{ Nucl. Phys.} B}
\def\PLA{{ Phys. Lett.}  A}
\def\PLB{{ Phys. Lett.}  B}
\def\PNAS{Proc. Natl. Am. Soc.}
\def\PRL{ Phys. Rev. Lett.}
\def\PRA{{ Phys. Rev.} A}
\def\PRE{{ Phys. Rev.} E}
\def\PRB{{ Phys. Rev.} B}
\def\PD{{ Physica} D}
\def\ZPC{{ Z. Phys.} C}
\def\RMP{ Rev. Mod. Phys.}
\def\EPJD{{ Eur. Phys. J.} D}
\def\SAB{ Sens. Act. B}
\title{
The fundamental unit of  quantum conductance and quantum diffusion for a gas of massive particles}

\author{Lino Reggiani}
\email{lino.reggiani@unisalento.it}
\affiliation{Dipartimento di Matematica e Fisica, ``Ennio de Giorgi'',
Universit\`a del Salento, via Monteroni, I-73100 Lecce, Italy}
\affiliation{CNISM,  Via della Vasca Navale, 84 - 00146 Roma, Italy}

\author{Eleonora Alfinito}
\affiliation{Dipartimento di Matematica e Fisica, ``Ennio de Giorgi'',
Universit\`a del Salento, via Monteroni, I-73100 Lecce, Italy}
\author{Federico Intini}
\affiliation {Department of Sciences and Methods for Engineering \\
Via Amendola 2, Pad. Morselli - 42122 Reggio Emilia, Italy
}
\date{\today}
\begin{abstract}
By analogy with the fundamental quantum units of electrical conductance 
$G_0^e=\frac{2 e^2}{h}$ and thermal conductance 
$K_0^t=\frac{2 K_B^2 T}{h}$
we define a fundamental  quantum unit of conductance, $G_0^m$, and  diffusion of a massive gas of atomic particles, respectively  given by
$$ G_0^m=\frac{m^2}{h}  \ , \ D_0=\frac{h}{m}$$
with $h$ the Planck constant, $K_B$ the Boltzmann constant, $T$ the absolute temperature, $e$ the unit charge and $m$ the mass of the atomic  gas particle that move balistically in a one dimensional medium of length $L$.  
The effect of scattering can be accounted for by introducing an appropriate transmission probability in analogy with the quantum  electrical conductance model introduced by Landauer in 1957.
For an electron gas  
$G_0^m=1.25 \times 10^{-27} \ Kg^2/(J s)$ and  
$D_0 = 7.3 \times 10^{-3}  \ m^2/s$, and
we found  a quantum expression for the  generalized Einstein relation that writes 
$$G_0^e = \frac{2e^2m}{h^2} D_0 $$

\end{abstract}
\maketitle 
%
This letter presents the definition of the universal quantum unit of conductance and diffusion coefficient for a massive gas thus completing the scheme of mesoscopic quantum transport coefficients already existing in the literature  \cite{greiner00}.
\par
By analogy with the electrical conductance, the classical definition of the  conductance for a classical gas of particles reads:
\begin{equation}
G^m = \frac{m N \tau} {L^2}  
=\frac{m^2 N} {{m \sqrt{\overline{v^{'2}_x} L}}}  \Gamma
\label{eq1a}
\end{equation}
where $m$ is the particle mass, $N$ the particles number, 
$\tau$ the scattering time, $\overline{v^{'2}_x}$ a mean squared differential
(with respect to carrier number) quadratic velocity component
along the $x$ direction \cite{gurevich79},  and $L$ the sample length.
\par
The second expression of Eq. (\ref{eq1a}) refers to a one dimensional sample of length $L$ 
with
$$
l=\sqrt{\overline{v_x^{'2}}} \tau
$$
the associated mean free path, and 
$$
\Gamma=\frac{l}{L}
$$
a transmission probability, that describes the collisions.
\par
The second form of Eq. (\ref{eq1a}) leads to  a quantized expression under the quantum condition
\begin{equation}
h = m  \sqrt{\overline{v_x^{'2}}} L
\label{eq2}
\end{equation}
for $h \geq m  \sqrt{\overline{v_x^{'2}}} L$, 
with $h$ the Planck constant.
\par
Indeed, by analogy with Landauer quantum-conductance model 
\cite{landauer57,buttiker86,imry99}, Eqs. (\ref{eq1a}, \ref{eq2})  define a quantum  conductance for a massive gas  satisfying a Landauer paradigm: conductance  is transmission:
\begin{equation}
G^m=\frac{Nm^2}{h} \Gamma
\end{equation}
that for the balistic condition, $\Gamma=1$ and $N=1$, gives the fundamental unit of  conductance for an atomic massive gas with elementary mass $m$.
\begin{equation}
G_0^m=\frac{m^2}{h} 
\end{equation}
For an electron gas  
$G_0^m=1.25 \times 10^{-27} \ Kg^2/(J s)$.
\par
By going  to diffusion, the classical definition of the 
longitudinal diffusion coefficient reads
 \cite{gurevich79,reggiani16,reggiani18}:
\begin{equation}
D_x=\overline{v^{'2}_x} \tau 
= \sqrt{\overline{v^{'2}_x}}{L} \Gamma
\label{eq1}
\end{equation}
\par
The second form of Eq. (\ref{eq1}) leads to  a quantized expression under the analog quantum condition in Eq. (\ref{eq2}).
\par
Even for diffusion, by analogy with Landauer quantum-conductance model 
\cite{landauer57,buttiker86,imry99}, Eqs. (\ref{eq1}, \ref{eq2})  define the fundamental unit of  quantum diffusion satisfying a Landauer paradigm: diffusion is transmission:
\begin{equation}
 D=\frac{h}{m} \Gamma
\end{equation}
that for the balistic condition, $\Gamma=1$, gives the fundamental
unity of diffusion for an atomic massive gas with elementary mass $m$.
\begin{equation}
D_0 = \frac{h}{m}
\end{equation}
For an electron gas $D = 7.3 \times 10^{-5} \ m^2/ s$ that should be compared with the experimental values in Si at $77 \ K$ of $6.0 \times 10^{-3}\ m^2/ s$  and of $1.6 \times 10^{-2}\ m^2/ s$  for electrons and holes, respectively. 
We remark that the two values of diffusion give a ratio  of $2.7$ in close agreement with the value of the corresponding effective mass of
$m_h/m_e = 0.53/0.19=2.8 $ \cite{reggiani85}.
The sample dimensions and the temperature in experiments were not consistent with the conditions posed by quantization, which explain the significant higher values of the experiments when compared  with  the theoretical quantum expectations.  
\par                                
Interesting enough, the single particle mass satisfies the kinetic definition
\begin{equation}
m = G_d \times D_0
\end{equation}
that is valid in both classical and quantum cases.
\par
In addition, the classical expression of the  generalized Einstein relation \cite{einstein05}
\begin{equation}
G_e =  \frac{e^2 m \overline{N}}{(Lm)^2 \overline{v_x}^{'2}} D 
\end{equation}
with $G_e$ the electrical conductance, $e$ the electrical charge unit,  and $\overline{N_e}$ the average total number of charge particles inside the sample, for a one dimensional geometry  takes the quantum form:  
\begin{equation}
 G_e =  \frac{2e^2 m}{h^2} D_0 
\end{equation}
We remark, that going from classical to quantum the average number of charge carriers is substituted by the number 2 relating to the number of transverse modes accounting for spin degeneracy.
We also notice that by considering an alternative Einstein relation of the form  $G_e \times D_0$, we obtain:
\begin{equation}
 G_e \times D_0  =  \frac{2e^2}{m} 
\end{equation}
that is fully compatible with the classical expression, in other words quantum effects are washed out. 
\par
The above results need some comments concerning the one dimensional and balistic conditions necessary to find the fundamental quantum units of the
kinetic coefficient considered, as for the other coefficients available  in  the literature \cite{greiner00}.
The above conditions give the macroscopic values in the form of  global quantities, that cannot be factorized in a geometrical and local contribution (for example the local conductivities) as for the case of the classical 3D case. The quantum results contains two basic quantities: the Planck constant as signature  of quantum mechanics, and a physical quantity that is the signature of the physical magnitude of interest.  In the present case the mass of the particles.     
 For the electrical and thermal conductance we had the electrical charge 
(or eventually a multiple of the unit in case of multiple-charged 
particles) and the Boltzmann constant coupled with the temperature in the case of thermal conductance. In a global representation, there is no connection between the external perturbation and the response as typical of a kinetic approach. In particular, the kinetic approach relates the kinetic coefficient basically to local scattering mechanisms, parametrized by a relaxation time $\tau$ in the simplest case. In the present balistic approach the 
kinetic coefficients are controlled by the contacts and the  external  reservoirs. For example, in our case balistic  diffusion implies a closed system. Accordingly, the number of particles inside the sample are constant in time. Thus, the only scatterings are internal specular reflection of particles at the opposite contacts. This reflection process is then responsible for the smoothing of any initial 
concentration gradient towards a final homogenous condition in the long time limit as expected by thermal equilibrium conditions.
By contrast,  balistic conductance implies  an open system so that the particle number inside the sample, being controlled by a chemical potential, is not constant in time. Thus, the contacts are perfectly transparent for particles going into and out from the sample, and the stochastic mechanism comes from the fluctuations of the total number of particle inside the sample that are responsible of the incoherent mechanism of entry and leaving of particles from the opposite contacts.  Thus, conductance can be determined  by making use  of the fluctuation-dissipation theorem 
already under thermal equilibrium conditions, that is in the absence of external forces \cite{reggiani16}. By considering a transmission probability, that can be less than unity, the presence of local scattering can be accounted for, as originally suggested by Landauer in 1957.
However, we want to stress that local scattering in the present model are not necessary to define the diffusion and conductance, and their inclusion 
has the net effect to  decrease the value of the fundamental unit dawn to a zero  value for a vanishing value of the transmission probability.
\par
In conclusion, by introducing a quantum unit of conductance and diffusion under 1D quantum balistic conditions we complete the 
quantum definition of four fundamental kinetic coefficients  of linear response \cite{greiner00}; i.e. conductance,  diffusion, electrical and thermal conductance. An alternative form of the  generalized Einstein relation evidences an intriguing property of being compatible with the classical result. 
We further notice that the classical definition of diffusion can be extended to relativistic particle, that is a  photon gas \cite{reggiani18} as:
\begin{equation}
 D_0 ^{rel} =  cL 
\end{equation}
with $c$ the light velocity in vacuum. 
Present findings concerning quantum conductance and diffusion are open to a further experimental validation.
\section{Acknowledgments}
Prof. Tilmann Kuhn from M\"unster University is warmly thanked for the very valuable comments he provided on the  subject. 

\begin{thebibliography} {99}
%

\bibitem{greiner00} 
A. Greiner,  L. Reggiani,  T. Kuhn  and  L. Varani,
Carrier kinetics from the diffusive to the ballistic regime: linear response
near thermodynamic equilibrium, {\it Semicond. Sci. Technol.} {\bf 15} (2000) 1071-1081.
%
\bibitem{gurevich79} 
S.  Gantsevich,  R.  Katilius and V.  Gurevich, 
Theory of fluctuations in nonequilibrium electron gas, {\it Rivista  Nuovo  Cimento} {\bf 2} (1979) 1-87.
%
\bibitem{landauer57}
R. Landauer, "Spatial Variation of Currents and Fields Due to Localized Scatterers in Metallic Conduction," in IBM Journal of Research and Development, vol. 1, no. 3, pp. 223-231, July 1957, doi: 10.1147/rd.13.0223.


\bibitem{buttiker86}
M  B\"uttiker,
Four-terminal phase-coherent conductance
Phys. Rev. Let. 57, 1751 (1986).

%
\bibitem{imry99}
Y. Imry and R. Landauer.
Conductance viewed as transmission
Rev. Mod. Phys. 71, S306 (1999).

\bibitem{reggiani16}
L. Reggiani, E. Alfinito and T. Kuhn, Duality and reciprocity of fluctuation-dissipation relations in conductors, Phys Rev.  94, 032112 (2016) 

\bibitem{reggiani18}
L. Reggiani, and E. Alfinito, 
Fluctuation dissipation theorem and electrical noise revisited, 
{\it Fluct. Noise Lett.}  {\bf 18} (2018) 1930001 (33 pages).
%
%
%
%
\bibitem{reggiani85}
 L. Reggiani,
Hot Electron Transport in Semiconductors, Springer Verlag Topics in 
Applied Physics, {\bf Vol. 58} (Berlin-Heidelberg, 1985).
%
\bibitem{einstein05}		
A. Einstein, 
Über die von der molekularkinetischen Theorie der Wärme geforderte Bewegung von in ruhenden Flüssigkeiten suspendierten Teilchen
(On the movement of small particles suspended in stationary liquids required by the molecular-kinetic theory of heat), 
{\it Annalen der Physik}  {\bf 322} (8) (1905) 549–560.	
%
%

\end{thebibliography}
\end{document}